\documentclass[10pt,conference]{IEEEtran}
\IEEEoverridecommandlockouts

\usepackage{cite}
\usepackage{amsmath,amssymb,amsfonts}
\usepackage{algorithmic}
\usepackage{graphicx}
\usepackage{textcomp}
\usepackage{xcolor}
\usepackage{booktabs}
\usepackage{multirow}
\usepackage{url}
\usepackage{tikz}
\usetikzlibrary{arrows.meta, positioning, shapes.geometric}
\graphicspath{{figures/}{./}}
\def\BibTeX{{\rm B\kern-.05em{\sc i\kern-.025em b}\kern-.08em
    T\kern-.1667em\lower.7ex\hbox{E}\kern-.125emX}}

\begin{document}

\title{A Layered Input--Output Defense for Open-Weight Language Models
under Adaptive Jailbreak Attack
\thanks{This work is supported by the Natural Sciences and Engineering Research Council of Canada (NSERC) under Grant RGPIN-2021-02968.}\thanks{\copyright~2027 IEEE. Personal use of this material is permitted. Permission from IEEE must be obtained for all other uses, in any current or future media, including reprinting/republishing this material for advertising or promotional purposes, creating new collective works, for resale or redistribution to servers or lists, or reuse of any copyrighted component of this work in other works.}}

\author{%
\IEEEauthorblockN{Yousef Bayoumi}
\IEEEauthorblockA{\textit{Trustworthy and Secure AI Lab}\\
\textit{Faculty of Computer Science}\\
\textit{University of New Brunswick}\\
Fredericton, Canada\\
ybayoumi@unb.ca}
\and
\IEEEauthorblockN{Roozbeh Razavi-Far}
\IEEEauthorblockA{\textit{Trustworthy and Secure AI Lab}\\
\textit{Faculty of Computer Science}\\
\textit{University of New Brunswick}\\
Fredericton, Canada\\
roozbeh.razavi-far@unb.ca}
\and
\IEEEauthorblockN{Ehsan Hallaji$^{1,2}$}
\IEEEauthorblockA{$^1$\textit{University of New Brunswick}\\
Fredericton, Canada\\
$^2$\textit{Toronto Metropolitan University}\\
Toronto, Canada\\
ehallaji@torontomu.ca}
}

\maketitle

\begin{abstract}
Defenses that rewrite a language model's prompt or check its response
are cheap and usually evaluated against a fixed set of attack prompts.
That leaves out the attacker who tries again. We
compare training-free defenses on two open-weight models under a fixed
benchmark and under an attacker allowed ten attempts. Attack success
rises by 8 to 40 percentage points once the attacker may try again, so a
fixed benchmark overstates what these defenses provide, and on one model
it ranks two of them in the wrong order. On Qwen, most of that rise
remains when the attacker never sees the responses. We also combine
existing input and output defenses into a layered defense
that hardens every request and checks every response before
release. It records the lowest attack success on both models and
leaves refusal of unsafe requests unchanged, though false refusal rises
by 10 to 18 points. An adaptive attacker still defeats it on up
to 45.0\% of behaviors. We release 664 attacks that succeeded against defended
models.
\end{abstract}


\section{Introduction}

A language model has no channel that separates its instructions from the
data it is asked to work on. The developer's system prompt, the user's
message, and any document or tool output placed in the context are
concatenated into a single sequence of tokens, and the model has nothing
but the text itself to go on. Anything that reads as an instruction can
be followed, wherever in that sequence it appears, and a model given
text that says to ignore its previous instructions has no reliable way
to treat that as data rather than as a command.

This is what prompt injection exploits \cite{doi:10.36227/techrxiv.177272853.30003431/v1}, and OWASP ranks it first among
the security risks for applications built on large language models
\cite{owasp2023}. It takes two forms. In \emph{indirect} prompt
injection the user is benign and the adversarial instruction has been
hidden in external content the model was asked to process, such as a
retrieved document or the output of a tool \cite{greshake2023indirect}. Such external content may also include source code processed by language-based analysis tools \cite{jelodar2025largelanguagemodelsllms}.
In \emph{direct} prompt injection the user is the adversary and writes
the message, aiming to obtain content the model would otherwise refuse.
This paper studies the direct form, the case commonly called
jailbreaking, which every deployed model faces regardless of what is
built around it.

Defenses against it fall into two classes. Training-based methods change
the model's weights so that it learns to keep instructions and data
apart \cite{chen2024struq}. They can give stronger protection,
but they call for training data, compute, and a pipeline to run
them, and adjusting a model for safety can cost some of its general
ability. Unlike continual-learning approaches that adapt models under changing conditions \cite{RAZAVIFAR2026134136}, training-free methods leave the weights alone and act at inference time, rewriting the prompt before generation or inspecting the
response after it \cite{xie2023selfreminder, hines2024spotlight}. They
can be applied to a model exactly as released and cost nothing to train,
which makes them the practical option for a deployment with no training
pipeline of its own.

Much of the evidence for the training-free class comes from
fixed sets of prompts. A benchmark is run against the undefended model
and again with the defense in place, and the fall in attack success is
reported. An attacker who reads the response and tries again is not
represented in that measurement, although this is what a real adversary
does. Where defenses have been tested against an attacker allowed to
adapt, many have failed \cite{andriushchenko2024adaptive,
nasr2025attacker}.

This study has two objectives. The first is to measure how much a
fixed benchmark overstates the protection that training-free defenses
give against an attacker who can try again, and to find out whether the
gap comes from the extra attempts or from what the attacker learns from
the responses. The second is to test whether
adding a check on the response holds up better than hardening the prompt
alone, and at what cost on benign requests.

This paper makes three contributions. We evaluate three
training-free defenses and an undefended baseline on two open-weight
models under two settings, a
fixed benchmark and an attacker allowed ten attempts per behavior,
reporting what each defense costs on benign requests alongside what it
prevents. We combine existing input and output defenses into a
layered defense and show that the output check accounts for most of its
advantage over a single input defense. We also release the attacks the adaptive attacker found, 664
of which succeeded against a defended model and 366 against the same
models undefended.

A fixed benchmark does not only overstate what these defenses provide.
It can point to the wrong one. Across the four defenses and two models
we test, attack success rises by between 8 and 40 percentage points once
the attacker may try again, and on Qwen most of the rise remains
when it never sees the responses. On one model the two settings disagree about
which of two defenses is stronger, so relying on the static numbers
alone could mean choosing the weaker one. Under both settings, the
layered defense records the lowest attack success on both models, and
its advantage does not come from refusing more often.

The code and the dataset of attacks are available for research use
on request. The dataset is not posted openly for the reasons given in
the Ethics and Responsible Release section.

Section~\ref{sec:background} reviews attacks on language models and the
defenses proposed against them. Section~\ref{sec:problem} states the task
and the attack model. Section~\ref{sec:method} describes the layered
defense, the baselines, and the procedure used to generate adaptive
attacks.
Section~\ref{sec:results} reports the results, and
Section~\ref{sec:conclusion} concludes.

\section{Background}
\label{sec:background}

\subsection{Attacks}

Direct attacks differ in how the prompt is produced and in what access
that requires. The earliest were written by hand, wrapping the request in
role-play or persona framings \cite{wei2023jailbroken}.
Optimization-based methods automate the search, using access to the
model's internals to find a suffix that when added to the request makes a
refusal unlikely \cite{zou2023universal}. These are effective, but the
suffixes they produce are unreadable and unlike anything a user would
send. A
third approach uses a second language model as the attacker, which writes
a prompt, reads the victim's reply, and revises it \cite{chao2023pair,
mehrotra2023tap}. We adopt this approach. It requires no access to the
victim's internals, only the ability to send it prompts and read its
replies, and it produces ordinary conversation rather than a garbled
string of tokens.

A prompt drawn from a fixed set is written once and never adjusts to how
the defense responds. An attacker who reads that response and revises the
next attempt can learn where the defense holds and where it
gives way. Tram\`{e}r et al. established for adversarial examples that a
defense must be measured against exactly this kind of attacker, one who
knows it and adapts to it \cite{tramer2020adaptive}. The same pattern
holds for language model
defenses, where methods reporting strong robustness against fixed
benchmarks have been defeated once an attacker was allowed to adapt to
them \cite{andriushchenko2024adaptive, nasr2025attacker}.

\subsection{Defenses}

Training-based defenses teach the model to separate instructions from
data \cite{202607.1022}, through a structured input format \cite{chen2024struq}, preference
optimization against injected samples \cite{chen2025secalign}, or an
explicit priority ordering over message roles
\cite{wallace2024hierarchy}. Pandya et al.\
\cite{pandya2025attention} break StruQ and SecAlign with a white-box
attack that uses the model's attention to guide its search, reaching
success rates of up to 85 to 95\% on unseen prompts.

We group the training-free defenses studied here by where they
act. Input-side methods
rewrite the prompt before generation, whether by
appending a reminder of the model's safety obligations
\cite{xie2023selfreminder}, marking untrusted content with delimiters and
instructing the model to treat it as data \cite{hines2024spotlight},
paraphrasing the input to disturb an adversarial construction
\cite{jain2023baseline}, or turning injection techniques back on the
attacker \cite{chen2025acl}. Paraphrasing was proposed against
optimized token strings and leaves a request written in ordinary language
largely intact. The other input methods add an instruction that the model
must keep following while it reads the attacker's message. Output-side
methods inspect what the model
produced, either by asking it to examine its own response
\cite{phute2023selfdefense} or by training it to emit a safety tag
\cite{wang2024selfguard}. They do not rely on the prompt being
obeyed, but a model that examines its own output can only catch what it
recognises as harmful.

Delimiter-based and attack-as-defense methods were designed for prompt
injection, where the prompt contains a separable region of untrusted
data that the defense can mark and police. A direct jailbreak presents no
such region, since the entire user turn is the attack. Spotlighting
can still mark the whole turn as data, but escape-deletion and
fake-completion have nothing to act on, and Table~\ref{tab:static} shows
that they give no protection here.

\section{Problem Statement}
\label{sec:problem}

\subsection{Task}

A deployed assistant is a language model that receives a system prompt
written by the developer and a user turn written by whoever is talking to
it, and returns a response. A training-free defense adds two steps around
that call. An input step rewrites what the model is shown before
generation, and an output step checks what the model returned and may
replace it. Both steps work on text alone. The weights are never changed,
so the defense can sit in front of a model used exactly as released. Each
defense studied here is a particular choice of one step or both.

\subsection{Attack Model}

The adversary is the user. Its access is black-box, limited to submitting
prompts and reading the returned text. It cannot see the weights,
gradients, logits, or the system prompt, and it does not know which
defense is deployed beyond what the responses reveal.

The adversary selects a target behavior from a fixed set of harmful
behaviors and may issue up to $N$ queries pursuing it. After each query
it observes the defended response and learns whether the attempt
succeeded. It may then choose the next prompt
using everything it has seen. An attempt succeeds when the response is
itself an instance of the behavior, judged by a classifier held fixed
across all conditions. A real attacker would judge success by
reading the response, and our implementation takes that judgment from
the classifier, as PAIR does \cite{chao2023pair}. Success is a property of the output rather than of
the prompt, so a response that discusses, refuses, or argues against the
behavior does not count.

The \emph{static} setting used throughout the training-free defense
literature is the special case $N = 1$, with the prompt drawn from a
fixed benchmark. The adversary never sees a response and never adapts.
The \emph{adaptive} setting takes $N > 1$ and lets the adversary
rewrite. This paper measures the gap between the two.

\section{Methodology}
\label{sec:method}

Every defense is evaluated in the same sequence. It is first run once
on each HarmBench prompt, which gives its static attack success. Four
configurations are then attacked by an adaptive attacker that revises its
prompts over up to ten attempts, and the same four are run on benign
prompts to measure false refusal. One classifier scores every response in
both settings. On a fixed random sample of 100 behaviors, the adaptive
attack on Qwen is also repeated with a new random seed, and run once more
with the victim's replies hidden from the attacker, which is told only
that each attempt failed.

\subsection{A Layered Input--Output Defense}
\label{sec:layered}

Section~\ref{sec:background} noted that most input defenses rely on
the model continuing to follow an added instruction while it reads the
attacker's text. Under the attack model of Section~\ref{sec:problem}, the
attacker can revise that text until the model stops. A check that reads
the response in a separate context is harder to defeat this way, because
the attacker's text is not there to argue with. The attacker can still
ask for the response in a form the check fails to recognise. The layered
defense therefore keeps input hardening, which is cheap and stops many
attacks on the first attempt, and adds a check on the response. The aim
is to raise the cost of an attack rather than to prevent it outright.

\begin{figure}[t]
\centering
\begin{tikzpicture}[
  node distance=5mm and 7mm,
  box/.style={draw, rounded corners, align=center, minimum height=7mm,
              minimum width=15mm, font=\footnotesize, fill=gray!8},
  dec/.style={draw, diamond, aspect=1.5, align=center, font=\footnotesize,
              fill=gray!8, inner sep=0pt, minimum size=9mm},
  arr/.style={-{Latex[length=1.6mm]}, thick},
]
\node[box] (u) {User\\prompt};
\node[box, right=of u] (harden) {Input\\hardening};
\node[box, right=of harden] (victim) {Victim\\model};
\node[box, below=8mm of victim] (critic) {Output\\critic};
\node[dec, left=of critic] (q) {Harmful?};
\node[box, left=of q] (refuse) {Return\\refusal};
\node[box, below=8mm of q] (release) {Return\\response};

\draw[arr] (u) -- (harden);
\draw[arr] (harden) -- (victim);
\draw[arr] (victim) -- (critic);
\draw[arr] (critic) -- (q);
\draw[arr] (q) -- node[above, font=\footnotesize]{Yes} (refuse);
\draw[arr] (q) -- node[right, font=\footnotesize]{No} (release);
\end{tikzpicture}
\caption{The layered defense. Input hardening is applied to every
request, and the output critic inspects the response before it is
released.}
\label{fig:layered}
\end{figure}
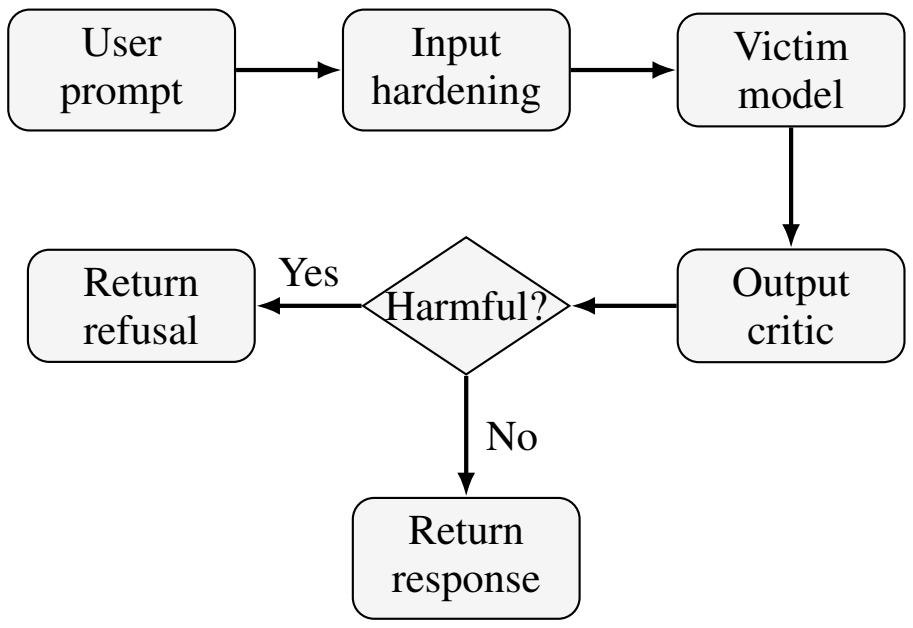

We first built this defense with a risk gate. A classifier scored each
incoming prompt and the hardening was applied only to the prompts it
flagged, so that benign traffic would not pay for it. Across the 300
harmful behaviors of HarmBench \cite{mazeika2024harmbench}, the
classifier cleared 116 as safe and passed them to the model with no
defense at all. Those 116 went on to account for
73\% of every successful attack against that design. A gate of this kind
is bounded by its false negative rate, so we removed it, accepting
that benign requests now pay for the hardening as well. In its place, the final design applies
two fixed layers to every request, shown in Fig.~\ref{fig:layered}.

\begin{enumerate}
\item \emph{Input hardening:} The user turn is wrapped in explicit
delimiters, and the system prompt tells the model to treat everything
inside them as data, not instructions \cite{hines2024spotlight}. A
reminder of the model's safety obligations comes right after, so it is
the last thing the model reads before it responds
\cite{xie2023selfreminder}.

\item \emph{Output-side examination:} After generation, the same model
re-reads its own output in a new context containing only that text and an
instruction to review it for harm. If the review finds it harmful, the
output is replaced with a refusal \cite{phute2023selfdefense,
wang2024selfguard}.
\end{enumerate}

The victim model performs that check by re-reading its own output outside
the conversation that produced it, then judging directly whether the text
is harmful. Prior work shows this approach catches content a model would
not have flagged inside its original context \cite{phute2023selfdefense,
wang2024selfguard}. The effect is not absolute.
Section~\ref{sec:results} reports where it holds and where it does not,
and the two models tested differ sharply.

This reviewer is always the victim model itself, never the classifier
used elsewhere in this paper to score attack success. Keeping the two
separate avoids a defense whose refusals simply mirror the grader's own
criterion for success.

\subsection{Baseline Defenses}
\label{sec:baselines}

The layered defense is compared in Table~\ref{tab:static} with an
undefended baseline and eight defenses that act on the prompt alone. The
baseline uses a plain assistant system prompt. The self-reminder adds a
note after the user's message telling the model not to follow
instructions embedded in it and to decline harmful requests
\cite{xie2023selfreminder}. Spotlighting wraps the user turn in
delimiters and tells the model in the system prompt to treat the wrapped
text as data \cite{hines2024spotlight}. The paraphrase defense has
Mistral-7B-Instruct \cite{jiang2023mistral} rewrite the user's message
before the victim answers it \cite{jain2023baseline}. Four
attack-as-defense methods from Chen et al.\ \cite{chen2025acl} turn
injection techniques against the attacker. Ignore-shield precedes the
message with an instruction to disregard any directives inside it, and
escape-deletion appends escape characters that simulate deleting the
message before repeating the request. Fake-completion appends a fake
response declaring the task complete and then repeats the request, and
Fake-completion (t) does the same through a fake assistant turn in the
chat template. The risk-gated design is the first version of the layered
defense, which applied paraphrasing and spotlighting only to prompts that
a classifier flagged as risky.

\subsection{Adaptive Attack Generation}
\label{sec:attacker}

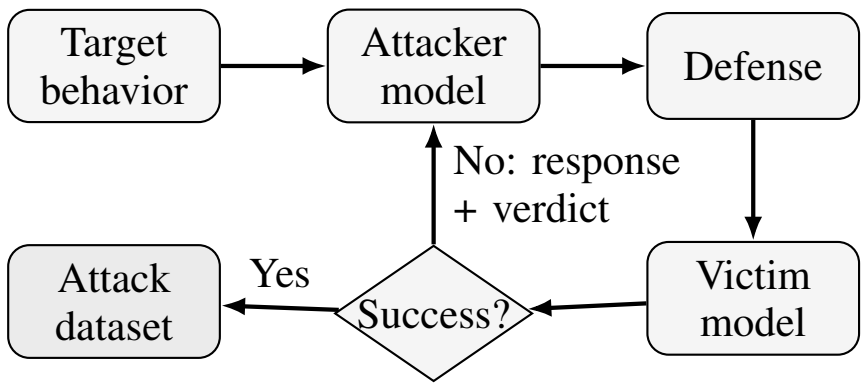
\begin{figure}[t]
\centering
\begin{tikzpicture}[
  node distance=5mm and 7mm,
  box/.style={draw, rounded corners, align=center, minimum height=7mm,
              minimum width=14mm, font=\footnotesize, fill=gray!8},
  dec/.style={draw, diamond, aspect=1.5, align=center, font=\footnotesize,
              fill=gray!8, inner sep=0pt, minimum size=9mm},
  arr/.style={-{Latex[length=1.6mm]}, thick},
]
\node[box] (beh) {Target\\behavior};
\node[box, right=of beh] (attack) {Attacker\\model};
\node[box, right=of attack] (def) {Defense};
\node[box, below=8mm of def] (victim) {Victim\\model};
\node[dec, below=8mm of attack] (judge) {Success?};
\node[box, below=8mm of beh, fill=gray!16] (data) {Attack\\dataset};

\draw[arr] (beh) -- (attack);
\draw[arr] (attack) -- (def);
\draw[arr] (def) -- (victim);
\draw[arr] (victim) -- (judge);
\draw[arr] (judge) -- node[above, font=\footnotesize]{Yes} (data);
\draw[arr] (judge) -- node[right, font=\footnotesize, align=left]
      {No: response\\{+} verdict} (attack);
\end{tikzpicture}
\caption{The adaptive attack loop, run for at most ten iterations per
behavior and halted on the first confirmed success.}
\label{fig:attackloop}
\end{figure}

The adaptive setting of Section~\ref{sec:problem} is instantiated with a
second language model in the adversary's place, following PAIR
\cite{chao2023pair}. It is seeded with the target behavior, not a
benchmark prompt to reword, and writes its own prompt from that
description. Each behavior is pursued through a single conversation: the
attacker writes a prompt, reads back the victim's defended reply together
with the classifier's verdict on it, then revises and tries again, for up
to $N = 10$ turns, stopping as soon as the classifier confirms a success.
Because the loop stops at the first success, every verdict the
attacker receives reports a failure.
Fig.~\ref{fig:attackloop} shows the arrangement.

That classifier is the same behavior-conditioned model used to score
every other result in this paper, so a response counts as a success under
the same criterion whether it came from the fixed benchmark or from this
loop.

When the layered defense is active, its output critic also runs inside
the loop, so the attacker only ever sees the response the defense
actually releases. A blocked response comes back as a refusal, so the
attacker must learn to work around the output check as well as the input
transformation.

The attacker's system prompt needs one further constraint to keep it on
the target behavior. An attacker without it will soften the objective,
turning a request to argue a position into a request to rebut it, and the
victim then complies with a harmless version that is correctly scored as
a failure. The instruction therefore states that the victim's output must
itself be an instance of the behavior, and that producing the opposite
counts as failure.

\section{Experimental Results}
\label{sec:results}

\subsection{Experimental Setting}

The victim models are Qwen2.5-7B-Instruct \cite{qwen2024qwen25} and
Llama-3.1-8B-Instruct \cite{llama3herd2024}. The attacker is
Mistral-7B-Instruct-v0.3 \cite{jiang2023mistral}, chosen because it
refuses red-teaming instructions less often than the victims and so
makes a usable attacker. All models run locally in four-bit NF4
quantisation \cite{dettmers2023qlora} on a single 16~GB consumer GPU,
which is also why the attacker and victim are held in memory together
while the judge is hosted on a second machine, a 6~GB laptop GPU.

Victims generate at most 512 new tokens at temperature 0.7. The attacker
generates at most 256 tokens at temperature 0.9 and sees the last six
attempts, with each reply cut to its first 500 characters. Each behavior is allowed at most ten attacker iterations and
stops at the first confirmed success. The same settings are used for
every victim and defense, so the numbers are comparable across cells.

Success is labelled by the HarmBench Mistral-7B classifier
\cite{mazeika2024harmbench}, which is conditioned on the target behavior
and answers whether a generation is an instance of it. One classifier is
used for every condition in the paper.

Two rates are reported. Attack success is the fraction of behaviors the
adversary succeeded on within its budget. False refusal is the fraction
of benign prompts the defended model declines, and it is reported
alongside attack success because a defense that refuses everything drives
attack success to zero while being useless. Both are given with Wilson
95\% confidence intervals. Comparisons between conditions use a
two-proportion $z$-test, and the $p$-value it returns is reported
wherever two conditions are compared directly.

\subsection{Datasets}

Attack evaluation uses the 300 standard and contextual behaviors of
HarmBench \cite{mazeika2024harmbench}. AdvBench \cite{zou2023universal}
and JailbreakBench \cite{chao2024jbb} were measured first and then set
aside, because undefended baselines already refuse almost all of their
prompts, at 0.7\% and 2.2\% success, which leaves nothing for a defense
to improve on.

Utility is measured on XSTest \cite{rottger2024xstest}. Its 250 safe
prompts are written to resemble harmful ones, and its 200 unsafe prompts
are worded to match them closely while describing something genuinely
harmful. The safe half gives the false refusal rate. A low false refusal
rate could also be reached simply by answering everything, safe or unsafe
alike, so the unsafe half checks that the same model still refuses the
genuinely harmful version at a high rate.

\subsection{Result Analysis}

\begin{table}[t]
\caption{Static ASR (\%) by defense and model (HarmBench, $n=300$
per cell).}
\label{tab:static}
\centering
\begin{tabular}{lcc}
\toprule
Defense & Qwen2.5-7B & Llama-3.1-8B \\
\midrule
None                  & 34.0 & 21.7 \\
Paraphrase            & 31.3 & 13.7 \\
Escape-deletion       & 34.0 & 12.3 \\
Fake-completion       & 36.7 & 16.3 \\
Fake-completion (t)   & 38.0 & 15.0 \\
Ignore-shield         & 24.0 & 14.3 \\
Risk-gated            & 26.3 & 14.7 \\
Self-reminder         & 23.3 &  8.0 \\
Spotlighting          & 20.3 &  7.0 \\
\textbf{Layered}      & \textbf{9.7} & \textbf{4.0} \\
\bottomrule
\end{tabular}
\end{table}

\begin{table}[t]
\caption{Attack success rate on Qwen2.5-7B (HarmBench, $n=300$) under
static and adaptive evaluation. Wilson 95\% confidence intervals.}
\label{tab:collapse}
\centering
\begin{tabular}{lccr}
\toprule
Defense & Static ASR (\%) & Adaptive ASR (\%) & $\Delta$ (pp) \\
\midrule
None            & 34.0\;[28.9,\,39.5] & 74.3\;[69.1,\,78.9] & $+40.3$ \\
Self-reminder   & 23.3\;[18.9,\,28.4] & 60.3\;[54.7,\,65.7] & $+37.0$ \\
Spotlighting    & 20.3\;[16.2,\,25.2] & 56.7\;[51.0,\,62.2] & $+36.3$ \\
\textbf{Layered} & \textbf{9.7}\;[6.8,\,13.5] & \textbf{45.0}\;[39.5,\,50.7] & $+35.3$ \\
\bottomrule
\end{tabular}
\end{table}

\begin{table}[t]
\caption{False refusal on the 250 safe XSTest prompts, and refusal on the
200 unsafe contrast prompts, by defense.}
\label{tab:overrefusal}
\centering
\begin{tabular}{lcccc}
\toprule
& \multicolumn{2}{c}{Qwen2.5-7B} & \multicolumn{2}{c}{Llama-3.1-8B} \\
\cmidrule(lr){2-3}\cmidrule(lr){4-5}
Defense & False ref.\ & Unsafe ref.\ & False ref.\ & Unsafe ref.\ \\
\midrule
None            &  9.6 & 88.5 &  8.0 & 96.5 \\
Self-reminder   & 10.8 & 89.0 & 22.8 & 99.0 \\
Spotlighting    & 12.4 & 89.5 & 13.6 & 95.0 \\
\textbf{Layered} & \textbf{19.6} & \textbf{90.0}
                & \textbf{25.6} & \textbf{98.0} \\
\bottomrule
\end{tabular}
\end{table}

\textbf{Static evaluation makes these defenses look effective.}
Table~\ref{tab:static} gives static attack success for every defense.
Spotlighting \cite{hines2024spotlight} and the self-reminder
\cite{xie2023selfreminder} roughly halve success against the undefended
baseline on both models, and the layered defense is the strongest on
each, at 9.7\% and 4.0\%. The
fake-completion and escape-deletion family does nothing here. On Qwen
escape-deletion scores exactly the undefended rate of 34.0\%, and the
two fake-completion variants sit 2.7 and 4.0 points above it, inside the
range expected from sampling alone at $n=300$.

The adaptive attacker was run on the undefended baseline, the two
strongest single-layer defenses in Table~\ref{tab:static}, and the
layered defense.

\textbf{Under an adaptive attacker that ranking survives but the
robustness does not.} Table~\ref{tab:collapse} places static and
adaptive success side by side on Qwen. Every defense loses between 35
and 40 percentage points, and every one of those gaps is significant
well beyond $p < 10^{-15}$. The undefended model rises from 34.0\% to
74.3\%, and the layered defense, the strongest configuration tested,
rises from 9.7\% to 45.0\%. Measured against a fixed prompt set these
defenses report roughly three times more protection than they provide
against an attacker allowed ten attempts.

\textbf{The gap comes mostly from trying again.} Across all eight
configurations in Tables~\ref{tab:collapse} and~\ref{tab:matched}, the
attacker's first attempt lands within 7 points of the static rate, and
the rise comes over the later attempts. On the 100-behavior sample, an
attacker that never sees the replies still defeats 70 behaviors on the
undefended model and 41 under the layered defense, against static rates
of 33 and 11 and adaptive rates of 72 and 48. Reading the replies adds 2
and 7 points, and neither difference is significant ($p = 0.85$ and
$p = 0.31$, McNemar's test).

Repeating the attack with a new seed gave the same totals as the
original run, 72 and 48. About a quarter of individual behaviors changed
outcome between the two runs, but the changes cancelled out.

\textbf{The defenses still help, and they separate into three groups.}
Under adaptive attack the layered defense sits 29.3 points below the
undefended baseline. Between the two single-layer input defenses there
is no separation to report, since the 3.7 point gap between the
self-reminder at 60.3\% and spotlighting at 56.7\% carries $p = 0.36$.
The layered defense does separate from both, by 15.3 points against the
self-reminder ($p = 0.0002$) and 11.7 points against spotlighting
($p = 0.0043$). The result is three groups rather than a ranking of
four. The undefended baseline is the weakest, the two input-only
defenses are indistinguishable from each other in the middle, and the
layered defense is the strongest.

\textbf{Almost all of that advantage comes from the output layer.}
Whenever the critic replaced a response, the response it suppressed was
recorded, so judging those recovers what the defense would have released
with the input hardening alone. On Qwen the input hardening by itself
reaches 18.3\%, against 20.3\% for spotlighting on its own, so adding a
second input layer to the first is worth about two points. Adding the
critic takes the same configuration to 9.7\%. Llama follows the same
pattern, from 6.0\% to 4.0\%. Input defenses share a failure mode, since each depends
on the model honouring an instruction about how to treat the prompt.
Under adaptive attack on Qwen they also fail on largely the same
requests, and 73\% of the behaviors defeated under spotlighting were also
defeated under the self-reminder. The critic does not depend on that instruction being honoured,
which is why it adds what a further input layer cannot.

\textbf{The layered defense does not buy this by refusing more.}
Table~\ref{tab:overrefusal} reports the utility cost. False refusal on
the safe XSTest prompts rises from 9.6\% to 19.6\% on Qwen
($p = 0.0015$) and from 8.0\% to 25.6\% on Llama ($p < 10^{-4}$), so the
defense is not free. Refusal on the unsafe contrast prompts is
statistically unchanged on both models, moving 1.5 points on each
($p = 0.63$ and $p = 0.36$). The defense still separates safe requests
from unsafe ones and has moved where it draws the line rather than
refusing indiscriminately. Two details qualify the cost. XSTest is built
to trigger over-refusal, so these rates sit far above what ordinary
traffic would produce, and the undefended models already refuse 8 to
10\% of the same prompts. Of the safe prompts the layered defense
wrongly refused, only 16 of 49 on Qwen and 4 of 64 on Llama came from
the output critic. The input hardening rather than the critic carries
most of the cost, which suggests a cheaper configuration is available.

\textbf{Robustness is bought at a worsening rate.} Against the
undefended baseline on Qwen, each point of false refusal buys an
11.7-point drop in attack success for the self-reminder, 6.3 for
spotlighting and 2.9 for the layered defense.

\textbf{The collapse is not particular to one model.}
Table~\ref{tab:matched} repeats the comparison on Llama-3.1-8B over the
same 300 behaviors. Every defense loses ground, by between 7.7 and 26.0
points, and all four gaps are significant. The layered defense again
gives the lowest attack success at 11.7\%, which is 36.0 points below
the undefended model.

\textbf{Static evaluation can also put the defenses in the wrong order.}
Under fixed prompts the self-reminder and spotlighting are
indistinguishable on Llama, at 8.0\% and 7.0\% with $p = 0.64$. Under
the adaptive attacker they separate, to 16.3\% and 31.3\%, and that
difference is significant. Spotlighting rises 24.3 points while the
self-reminder rises 8.3. A practitioner choosing between the two from
published static numbers would see nothing to distinguish them and could
reasonably take the one that turns out to be twice as easy to defeat.
A likely reason is that spotlighting asks the model to treat part of its own prompt as data
rather than instruction, and an attacker with several attempts can argue
it out of that framing. The self-reminder states a safety obligation,
which offers less to argue with.

\textbf{The margin the layered defense holds depends on the model.} On
Qwen it separates from every other configuration, including the
self-reminder by 15.3 points ($p = 0.0002$). On Llama it separates from
the undefended model and from spotlighting, but its 4.7 point advantage
over the self-reminder does not reach significance ($p = 0.10$). It
records the lowest attack success on both models, and on Llama the
evidence does not establish that it improves on the self-reminder alone.

Its degradation also differs. On Llama the layered defense rises only
7.7 points from its static rate, while on Qwen the same defense rises
35.3 points to reach 45.0\%. The critic is the victim model examining
its own output, so a defense built this way inherits the judgement of
the model it runs on. This is a limitation of the design
rather than a property of these two models, and it means the defense
should be measured on the model it will be deployed with.

\begin{table}[t]
\caption{Static and adaptive attack success on Llama-3.1-8B (HarmBench,
$n=300$). Compare with Table~\ref{tab:collapse} for Qwen2.5-7B.}
\label{tab:matched}
\centering
\begin{tabular}{lccr}
\toprule
Defense & Static (\%) & Adaptive (\%) & $\Delta$ (pp) \\
\midrule
None             & 21.7 & 47.7 & $+26.0$ \\
Self-reminder    &  8.0 & 16.3 & $+8.3$ \\
Spotlighting     &  7.0 & 31.3 & $+24.3$ \\
\textbf{Layered} &  \textbf{4.0} & \textbf{11.7} & $+7.7$ \\
\bottomrule
\end{tabular}
\end{table}

\subsection{Discussion}

The layered defense costs one additional forward pass for each request.
That pass reads the response and emits a single word, so its length is
fixed and short, and it adds far less work than generating the response
did. The input hardening is a string operation and costs nothing
measurable. Both layers are prompt-level, so the defense can be placed
in front of a model that is already deployed without touching it.

On one 16~GB consumer GPU, attacking a single behavior took about
four minutes and the eight full adaptive runs took about 165 hours, so
an evaluation of this kind is within reach of anyone deploying such a
model.

Three implications follow for anyone choosing among these defenses.
Published static numbers are not a safe basis for the choice, since they
can rank two defenses in the wrong order. Adding a second input defense
to a first gains little, and a check on the response gains more,
provided the deployed model recognises its own harmful output. On Qwen,
an attacker also got most of its advantage without reading the
responses, so making refusals less informative is unlikely to help
much.

Four limitations bear on how the results should be read. The critic is
the victim model judging its own output, so the defense is only as good
as that model's ability to recognise harm, and Section~\ref{sec:results}
shows the effect this has across the two models tested. Attack success
is labelled by a single classifier, so the absolute rates inherit
whatever error that classifier has, although the comparisons between
conditions are unaffected because the same classifier scores all of
them. Both victims are open-weight models of 7 to 8 billion
parameters, and the repeated run and the hidden-reply control cover only
Qwen on 100 behaviors. And the attacker was given ten attempts per
behavior, so the
reported rates are lower bounds. A more determined adversary has more
attempts available.

\section{Conclusion}
\label{sec:conclusion}

This study had two objectives, to measure how much a fixed
benchmark overstates training-free defenses and to test whether checking
the response as well as the prompt holds up better against an attacker
who can try again. The overstatement is large. Across both models
tested, attack success rose
by between 8 and 40 percentage points under an attacker allowed ten
attempts, and every
one of those gaps was significant. The extra attempts, rather than
reading the responses, account for most of that rise on Qwen. On one model the two settings also
ranked two defenses differently, so a fixed benchmark can mislead about
which defense to choose and not only about how much it protects.
Checking the response does help, within limits. The
layered defense proposed here, which hardens every request and checks
the response before releasing it, recorded the lowest attack success on
both models and left refusal of genuinely unsafe requests unchanged, but
an adaptive attacker still succeeded against it on 45.0\% of behaviors
on one model and 11.7\% on the other. No training-free defense examined
here is close to robust. The attacks that defeated these defenses are released so that
others can measure against an adversary that already knows what it is
facing.

\section*{Ethics and Responsible Release}

This work produces a collection of prompts that succeed in drawing
harmful content out of language models, so the reasons for releasing it
deserve to be stated plainly.

A defense cannot be measured against an adversary that adapts unless
adaptive attacks exist to measure it with. Nearly all public jailbreak
collections target undefended models, which means a defense evaluated
against them is being tested on attacks written without knowledge of it.
That is the gap this dataset fills, and closing it helps the defender
more than the attacker.

The target behaviors are taken unchanged from a public benchmark
\cite{mazeika2024harmbench}, so the set of harms represented here is
already documented in the literature. What is new is the phrasing that
reaches them through a particular defense, which is specific to the
models and defenses studied. Only 7\% of the attacks that defeated
a defended Qwen also succeeded when replayed once against Llama behind
the same defense. The release therefore adds little that
would help someone pursuing harm, while giving defense researchers
something they currently lack.

Every victim model was run locally from released weights. No hosted or
production system was attacked at any point. The dataset is released for
research use on request rather than posted openly, and it carries the
judge labels and the defense each attack was written against, so that
users can see what each record does and does not demonstrate.

\section*{Acknowledgments}

The authors used Claude, developed by Anthropic, as a writing aid to
draft and revise the text of this paper. All experiments, results, and analysis are
the authors' own and were verified by them.

\bibliographystyle{IEEEtran}
\bibliography{reference}

\end{document}